\documentclass[12pt]{iopart}
\usepackage{citesort}
\usepackage{graphicx}
\usepackage{color}
\usepackage{booktabs}
\usepackage{lscape}
\usepackage{multirow}
\usepackage[export]{adjustbox}
\usepackage[linesnumbered,lined,boxed,commentsnumbered,ruled]{algorithm2e}
\usepackage[noend]{algpseudocode}

\begin{document}

\noindent Accepted for publication in the Journal of Neural Engineering \\
Citation: Xilin Liu and Andrew G Richardson 2021 J. Neural Eng. 18 046034\\

\title[]{Edge Deep Learning for Neural Implants: A Case Study of Seizure Detection and Prediction}

\author{Xilin Liu and Andrew G. Richardson}

\address{University of Pennsylvania, Philadelphia, PA, USA}

\ead{xilinliu@seas.upenn.edu}
\vspace{10pt}

\begin{abstract}
\emph{Objective.} Implanted devices providing real-time neural activity classification and control are increasingly used to treat neurological disorders, such as epilepsy and Parkinson's disease. Classification performance is critical to identifying brain states appropriate for the therapeutic action (e.g. neural stimulation). However, advanced algorithms that have shown promise in offline studies, in particular deep learning (DL) methods, have not been deployed on resource-restrained neural implants. Here, we designed and optimized three DL models or edge deployment and evaluated their inference performance in a case study of seizure detection. \emph{Approach.} A deep neural network (DNN), a convolutional neural network (CNN), and a long short-term memory (LSTM) network were designed and trained with TensorFlow to classify ictal, preictal, and interictal phases from the CHB-MIT scalp EEG database. A sliding window based weighted majority voting (WMV) algorithm was developed to detect seizure events based on each DL model's classification results. After iterative model compression and coefficient quantization, the algorithms were deployed on a general-purpose, off-the-shelf microcontroller for real-time testing. Inference sensitivity, false positive rate (FPR), execution time, memory size, and power consumption were quantified. \emph{Main results.} For seizure event detection, the sensitivity and FPR for the DNN, CNN, and LSTM models were 87.36\%/0.169 $h^{-1}$, 96.70\%/0.102 $h^{-1}$, and 97.61\%/0.071 $h^{-1}$, respectively. Predicting seizures for early warnings was also feasible. The LSTM model achieved the best overall performance at the expense of the highest power. The DNN model achieved the shortest execution time. The CNN model showed advantages in balanced performance and power with minimum memory requirement. The implemented model compression and quantization achieved a significant saving of power and memory with an accuracy degradation of less than 0.5\%. \emph{Significance.} Inference with embedded DL models achieved performance comparable to many prior implementations that had no time or computational resource limitations. Generic microcontrollers can provide the required memory and computational resources, while model designs can be migrated to application-specific integrated circuits (ASICs) for further optimization and power saving. The results suggest that edge DL inference is a feasible option for future neural implants to improve classification performance and therapeutic outcomes.

\vspace{2pc}



\end{abstract}

\section{Introduction}

Many brain injuries and diseases may be treated by implanted devices that provide real-time classification of neural activity to produce a suitable control output. For example, closed-loop neuromodulatory devices control neural activity classified as pathological with electrical stimulation to treat epilepsy and Parkinson's disease \cite{sun2014closed}. Brain-machine interface (BMI) devices classify volitional intent to control external communication or movement devices for paralyzed individuals \cite{mak2009clinical}. In both cases, the neural implants record brain activity, select and extract relevant activity features, and perform classification and control on the basis of these features.

Performance of these devices is largely dependent on the feature selection and classification algorithms. Feature selection aims to transform the often noisy, correlated signals from many recording channels into a few non-redundant, informative inputs to the machine learning classifier. Feature selection is often a manually-specified, time-consuming process that requires substantial domain expertise \cite{lawhern2018eegnet}. Furthermore, current implantable devices such as the Neuropace RNS and Medtronic Activa PC+S closed-loop neurostimulators have restrained computation resources. Thus, only simple classification algorithms, such as feature thresholding or linear discriminant analysis, have been implemented \cite{baldassano2019cloud}.

Deep learning (DL) is a machine learning algorithm that has recently been applied to research-grade, non-implantable neural interface devices to improve performance \cite{roy2019deep}. Specifically, DL combines feature extraction, feature selection, and classification into a single framework, jointly optimizing the end-to-end process \cite{lecun2017deep}. When sufficient training data is available \cite{lotte2018review}, DL can achieve superior performance compared to more conventional algorithms \cite{schirrmeister2017deep}, especially in distinguishing hidden features critical for classification \cite{tsiouris2018long}. Furthermore, the DL approach is robust and more generalizable across different applications \cite{lawhern2018eegnet}. Although numerous DL approaches for neural interface devices have been studied, nearly all have done so offline \cite{lotte2018review}. While DL training is computationally intensive and likely to remain offline, DL inference must be performed online for real-time control. This is a particularly challenging problem for clinical-grade devices operating on battery-powered microprocessors or integrated circuits with limited computational resources and energy budget.

\begin{figure*}[tp]
  \centering
  \includegraphics[width=1\textwidth]{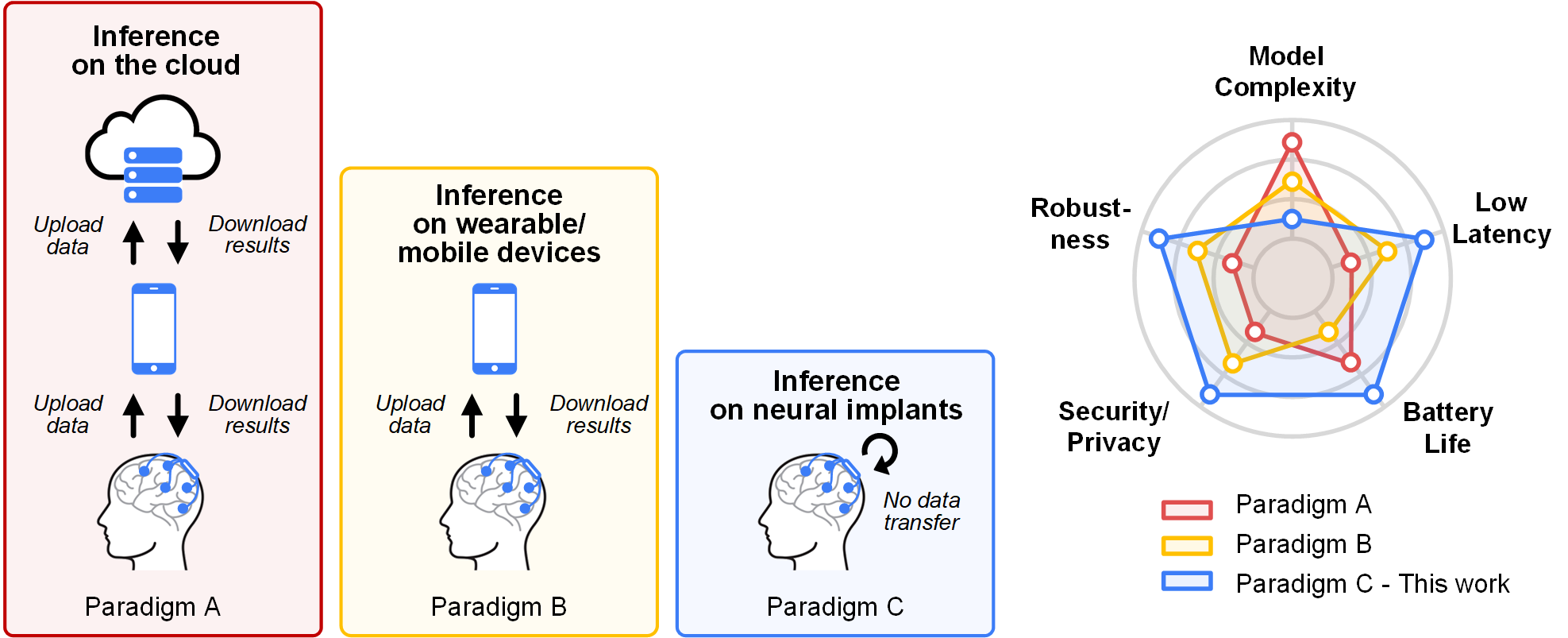}\\
  \caption{\emph{Left}, Illustration of three paradigms for executing deep learning inference. Paradigm A: DL inference on the cloud via internet or telecommunication. Paradigm B: DL inference on the wearable device via local wireless communication. Paradigm C: DL inference directly on the neural implants without any data transfer. \emph{Right}, The qualitative strengths and weaknesses of each paradigm.}\label{overview_inference}
\end{figure*}

To implement real-time DL inference for clinical neural implants, three paradigms have been proposed (Figure \ref{overview_inference}, left). First, inference can be done remotely through cloud computing (Paradigm A) \cite{baldassano2019cloud}. In this paradigm, the neural implant transmits the recorded data to a wearable device via local wireless communication, and the wearable device in turn uploads the data to a cloud-based workstation via internet or telecommunication. The cloud-based inference result is downloaded wirelessly back to the wearable device and then to the neural implant for producing the control output (e.g. neural stimulation). Second, inference can be done on the wearable device without the need for data transfer to a cloud-based platform (Paradigm B) \cite{mahmood2019fully}. Third, DL inference can be performed directly on the neural implant itself, requiring no data transfer to another device (Paradigm C) \cite{kiral2018epileptic}. Each of these three inference paradigms has its strengths and weaknesses (Figure \ref{overview_inference}, right). Although the cloud computing in Paradigm A offers the best possible inference accuracy, the latency and the robustness are the worst, which can adversely effect the timing-critical closed-loop therapeutic intervention. There are also concerns about cybersecurity and data privacy due to the required data transfer via internet or telecommunication \cite{naufel2020darpa}. Paradigm B eliminates the dependence on remote data transfer and the associated concerns, but it still relies on the robustness and security of the local wireless communication between the wearable devices and the neural implants. Moreover, the inference model complexity will be limited by the computational and power resources of the wearable devices. Paradigm C, using edge computing \cite{Hartmann2019edge}, avoids the disadvantages and concerns of wireless communication and can therefore potentially achieve the best robustness, the lowest latency, and the minimum security risks. Even if occasional wireless data transfer is performed in Paradigm C for offline analysis and performance assessments, the therapy itself would be robust to data transfer disruption and the opportunity for malicious attack would be the smallest of the three paradigms due to infrequency of transmission. However, a major question regarding Paradigm C is whether an edge DL model can achieve desired inference performance. Although machine learning-enabled processors for neural implants have been reported \cite{o2018nurip,zhu2020closedloop}, DL models have only rarely been implemented \cite{heller2018hardware,hugle2018early,kiral2018epileptic}. Therefore, the objective of the present work was to evaluate the inference performance limitations and resource constraints of edge DL designs through a case study.

For a DL model to be successfully deployed on a neural implant for real-time inference, the model design needs to meet three requirements. First, the model's inference performance (e.g. accuracy, sensitivity, specificity, etc.) should meet the requirements of the target application, including potential degradation from model compression and quantization. Second, the model's inference time should meet the latency requirement, especially in closed-loop applications. This not only requires a sufficient computational speed, but also limits the maximum length of the buffered input data segment, resulting in a constraint on the model's architecture. Third, the power and resource costs of the DL model should meet the budget of the neural implant. For instance, the available memory resources, both non-volatile storage memory and random access memory (RAM), limit the total number of trainable model parameters.

Potential hardware platforms for neural implants with edge DL include dedicated AI accelerators and general-purpose low-power processors. Dedicated AI accelerators provide higher energy efficiency than general-purpose processors. However, commercially available AI accelerators, such as IBM's TrueNorth \cite{akopyan2015truenorth}, Google's Edge TPU \cite{cass2019taking}, and Intel's Loihi \cite{davies2018loihi}, are too power hungry for neural implants. Also, large volume mobile devices-oriented AI intellectual property cores are not easily accessible for research-grade prototyping \cite{wang2019edge}. Ultra low-power AI accelerator chips have mainly been developed in research labs \cite{si2019twin,jia2020programmable,chen2019deep}. On the other hand, general-purpose low-power processors, such as ARM Cortex\textsuperscript{\textregistered}, Texas Instruments MSP430\textsuperscript{\textregistered}, and open-source reduced instruction set computer (RISC)-based microcontrollers (MCUs) provide a solution for low-cost, rapid prototyping for medical research and pre-clinical trials. If a further reduction in power or device footprint is desired, the MCU-based design can be migrated to an application-specific integrated circuit (ASIC) by integrating the MCU core together with analog neural recording and stimulation circuits, with an optional wireless communication module \cite{o2018nurip,liu2016design}. Thus, in this work, we focused on edge DL design using a general-purpose MCU.

To investigate DL design and optimization methods fulfilling practical requirements of neural implants, we conducted a case study of epileptic seizure detection. Real-time seizure detection and intervention through closed-loop neural stimulation has proven to be an effective treatment for medically-refractory epilepsy \cite{heck2014two}. Due to the importance of accurate, low-latency detection by a battery-powered neural implant, this application could benefit from edge computing. We adopted three commonly used DL architectures and customized the design for seizure detection on a MCU. We validated the methods using a publicly available annotated epilepsy database. The models were trained offline using Tensorflow. Compression and quantization methods were investigated for reducing the computational cost, memory, and power consumption for real-time inference. The inference results were compared to prior publications with no time and computational resource limitations. The strengths and weaknesses of each model were analyzed. Finally, we discuss the remaining challenges and future efforts toward implementing this novel paradigm in clinical neural implants.

\section{Methods}

\subsection{Data Preparation}
This study used the Boston Children's Hospital (CHB)-MIT scalp electroencephalography (EEG) database \cite{shoeb2009application}, which is publicly available at PhysioNet.org \cite{goldberger2000physiobank}. Although scalp EEG is inconsistent with signals utilized by real neural implants, the CHB-MIT database is one of the most popular epilepsy data sets used for benchmarking, allowing a fair performance comparison between our edge DL approach and prior methods. The database contains recordings collected from 23 pediatric patients with intractable epilepsy. The placement of the surface electrodes followed the international 10-20 system \cite{homan1987cerebral}. The original recordings were sampled at 256 Hz and 16-bit resolution, with a 60 Hz notch filter applied for removing the mains interference. The seizure start and stop time points for each patient were manually annotated by clinical experts. Channels that were not constantly available throughout the entire duration in each case were excluded in developing the algorithms, as suggested in \cite{tsiouris2018long}. No signal processing was used before training and testing the models.

Data segments during ictal, preictal, and interictal phases were selected for training the DL models. The ictal phases were well defined by the expert annotated seizure start and stop time points. There is no consistent definition of preictal phases in the literature, and the occurrence of preictal characteristics may differ across patients. In this work, we used the 3 min of recordings that are 30 sec prior to each seizure's start time as the preictal phases. It should be noted that preictal characteristics could happen much earlier than this period \cite{litt2002prediction}. Interictal segments were selected at least 2 hours before and after any ictal phase to avoid potential signal contamination \cite{truong2018convolutional}. An even longer interval between ictal and interictal segments up to 4 hours may further improve the classification performance \cite{tsiouris2018long,daoud2019efficient}. Unfortunately, the recordings of those phases are not always available in the chosen database. Postictal recordings were not used in this study.

The recordings from the interictal and preictal phases were much longer than the ictal phase. If all data segments were used in training, the severe imbalanced class distribution would cause a bias in the model's prediction \cite{japkowicz2002class}. We adopted two methods to address this issue. First, we generated more ictal segments by sliding the window with an overlap of 50\% \cite{truong2018convolutional}. Second, we applied a class weight to the loss function \cite{lawhern2018eegnet}. Interictal segments were randomly selected during day and night to avoid overfitting to irrelevant activities.

\subsection{Deep Learning Models}

\begin{figure}[!ht]
  \centering
  \includegraphics[width=0.7\textwidth]{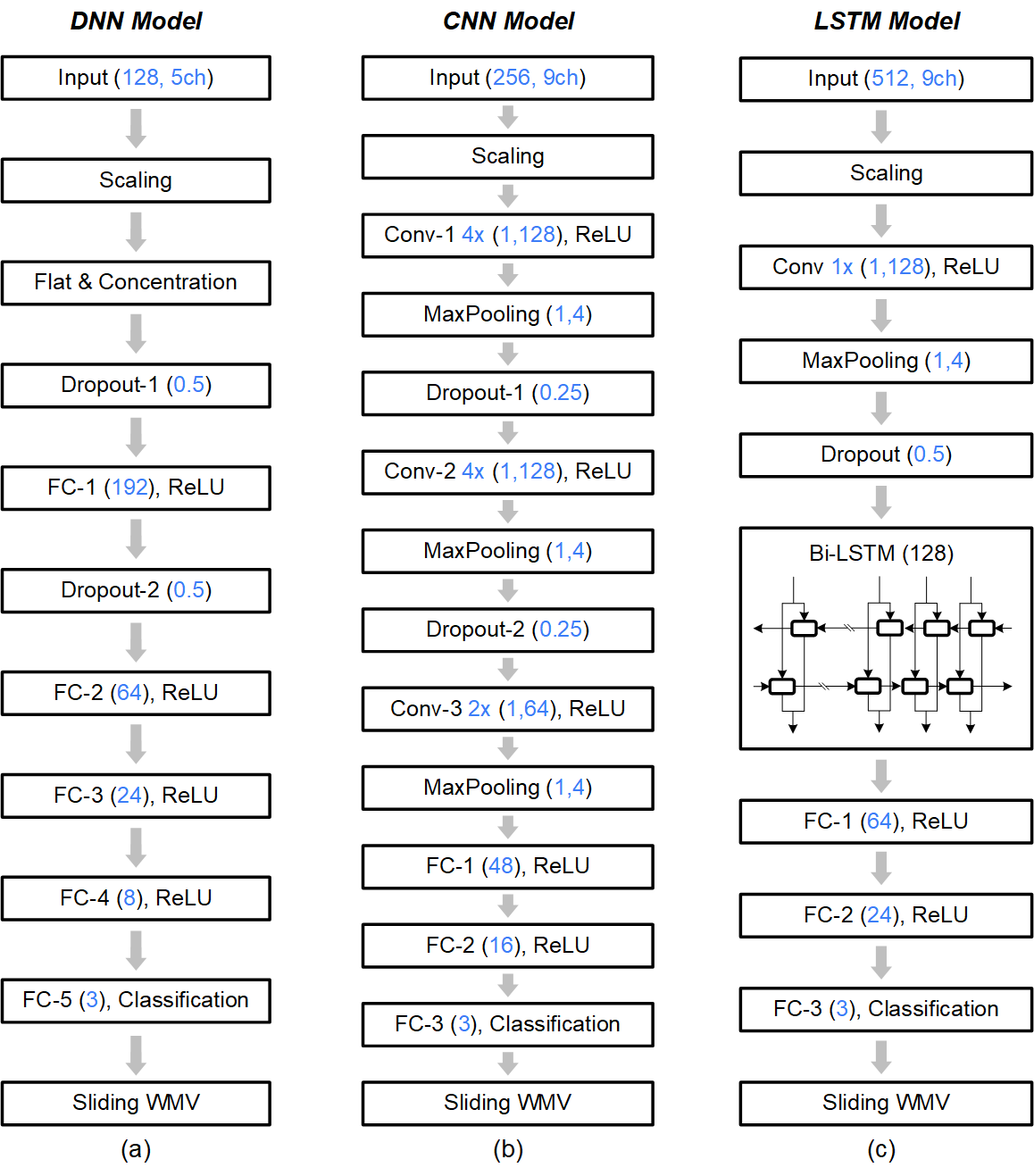}\\
  \caption{Simplified diagrams of the three DL models used in this study: (a) the deep neural network (DNN) model, (b) the convolutional neural network (CNN) model, and (c) the bidirectional long short-term memory (LSTM) model. Parameters of each layer are highlighted in blue.}\label{dl_models}
\end{figure}

We investigated three DL models for the seizure detection task: 1) a deep neural network (DNN) model, 2) a convolutional neural network (CNN) model, and 3) a bidirectional long short-term memory (LSTM) network model. The architectures of the three models are illustrated in Figure \ref{dl_models}. Although the model training was patient-specific, each model's architecture and parameter settings were kept the same for all patients in the database. The design details are presented in the following sections. 

\subsubsection{Deep Neural Network (DNN)}

DNN is a type of feedforward artificial neural network that consists of multiple layers. Each layer contains many processing units, namely artificial neurons. The processing function of each neuron is given in the Appendix. By connecting these neurons, DNNs emulate the way the brain processes information. Each neuron's output is processed by an activation function. In this work, the rectified linear unit (ReLU) function is used in all models \cite{nair2010rectified}. Conventional artificial neural networks often use nonlinear activation functions, such as hyperbolic tangent and sigmoid. However, these functions make gradient-based training challenging as the number of network layers increases, an issue known as the gradient vanishing problem \cite{goodfellow2016deep}. The introduction of the ReLU function overcomes this limitation by preserving the linear properties of positive values for gradient-based optimization, while still providing nonlinearity by setting all negative values to zero \cite{nair2010rectified}. Moreover, the ReLU function is computational friendly, which is important for the purposes of this work.

Our DNN model consists of a scaling layer, a flat and concatenation layer, followed by five layers of fully-connected (FC) neurons. The scaling layer removes the DC offset from each channel and linearly scales the input data based on the dynamic range. The FC layers form a pyramid shape with $N$ neurons in the first layer, where $N$ is the length of the input segment of one channel. A 50\% dropout layer is inserted before each of the first two FC layers. The class (ictal, preictal, or interictal) that has the highest activation function value is the final classification result.

\subsubsection{Convolutional Neural Network (CNN)}

CNNs have achieved extensive success in visual image recognition and natural language processing  \cite{krizhevsky2012imagenet,kim2014convolutional}. Using CNNs in seizure detection tasks has also been reported \cite{truong2018convolutional,khan2017focal,abiyev2020identification}. Convolutional (Conv) layers can be trained to automatically extract underlying features that best represent the data without human intervention. Given their superior ability in extracting features from images, past studies have converted time-domain EEG segments into spectrograms to be used as the CNN inputs \cite{truong2018convolutional}. However, the computational costs of the Fourier or wavelet transform of multiple EEG channels are prohibitive for real-time applications using low-power MCUs. In this work, we used Conv layers to extract features directly from time-domain EEG signals.

Instead of using 2-D standard Conv filters, we used 1-D Conv filters to process each EEG channel. This allowed us to reduce the model size and computational cost, while still preserving the most important features \cite{lawhern2018eegnet}. Our CNN model consists of three Conv layers, all using 'same' padding and stride=1. Each of the first two Conv layers consists of 4 kernels with a length of $fs$/2, where $fs$ is the sampling rate of the input EEG signal. The third Conv layer consists of 2 kernels with a length of $fs$/4. A max-pooling layer with a length of 4 was added to each of the Conv layers. The max pooling was used to prevent overfitting while reducing the computational costs. A 25\% dropout layer was inserted after each of the first two Conv layers to regularize the model. The outputs of the Conv layers were concatenated and processed by three FC layers.


\subsubsection{Long Short-term Memory Network (LSTM)}

LSTM is a type of recurrent neural network (RNN). In contrast to the feedforward neural networks described above, a RNN has recurrent connections that are suitable for capturing sequential information in the data. In particular, gating functions are used in each cell of the LSTM layers to control precisely what information is to be kept in the network and what is to be removed. Thus, LSTM has an inherent advantage in extracting certain temporal characteristics in time-domain signals \cite{gers1999learning}, which is crucial in tasks such as seizure detection.

In this work, we adopted a bidirectional LSTM network. Bidirectional LSTM networks process sequential information from two opposite directions simultaneously, which has proven useful in applications such as speech processing \cite{graves2013hybrid}. For the gate activation function, we used hard-sigmoid instead of sigmoid to avoid the exponential operation, and we used softsign as the state activation function. The processing and activation functions of the LSTM model are given in the Appendix.

We adopted a topology that combines an input Conv layer with the bidirectional LSTM layers. 
One kernel with a length of $fs$/2 was used in the Conv layer, followed by a max-pooling layer with a length of 4. A 50\% dropout layer was inserted between the input Conv layer and the bidirectional LSTM layers. Each of the bidirectional LSTM layers consists of 128 cells. Finally, FC layers convert the outputs for classification.

\subsubsection{Sliding Window Based Weighted Majority Voting}
Short data segment based classification often suffers from a trade-off between sensitivity and FPR \cite{baldassano2019cloud}. Since not every single data segment in the preictal phase exhibits signal characteristics that are related to an oncoming seizure, there is usually a limitation of the achievable classification accuracy before model overfitting. To achieve a high sensitivity while minimizing the FPR, we propose a novel sliding window based weighted majority voting (WMV) algorithm. The algorithm was implemented for real-time operation, as described in the pseudo-code of Algorithm \ref{alg_voting}.
\begin{algorithm}[!ht]
\small
\caption{Sliding window based WMV algorithm}\label{alg_voting}
 \SetKwInOut{Input}{Inputs}\SetKwInOut{Output}{Output}
 \Input{$Pred\_Seg\,[\,1:M\,] \in \{Ictal, \; Preictal, \; Interictal \}$, \; \textit{M: total segments in the win}, \\$Parameters: \alpha_{I, P}, \; \beta_{I, P}, \; and \;\; \theta_{I, P}$}
 \Output{$Pred\_Event \in \{Ictal, \; Preictal, \; Interictal \}$}
\BlankLine
\emph{\textbf{Initialization}:\\
$Score\,[\,Ictal\,] \leftarrow 0 $, $Score\,[\,Preictal\,] \leftarrow 0$, $Score\,[\,Interictal\,] \leftarrow 0 $ \\ 
$Acc\,[\,Ictal\,] \leftarrow 0 $, $Acc\,[\,Preictal\,] \leftarrow 0$ \\ $Pred\_Event \leftarrow Interictal$
}\;
\BlankLine
 \For{$i\leftarrow 1$ \KwTo $M$}{
    \Switch{based on {\textup{Pred$\_$Seg\,[\,i\,]}}}{
        \Case{Ictal}{
            $Score\,[\,Ictal\,] \leftarrow$ += $\alpha_I + \beta_I \cdot Acc\,[\,Ictal\,]$ \\
            $Acc\,[\,Ictal\,] \leftarrow $ += $1$,
            $Acc\,[\,Preictal\,] \leftarrow 0$
        }
        \Case{Preictal}{
            $Score\,[\,Preictal\,] \leftarrow$ += $\alpha_P + \beta_P \cdot Acc\,[\,Preictal\,]$ \\
            $Acc\,[\,Preictal\,] \leftarrow $ += $1$,
            $Acc\,[\,Ictal\,] \leftarrow 0$
        }
        \Case{Interictal}{
            $Acc\,[\,Ictal\,] \leftarrow 0$, $Acc\,[\,Preictal\,] \leftarrow 0$
        }
    }
     \BlankLine
     \uIf{$Score\,[\,Ictal\,] > \theta_I$}{$Pred\_Event \leftarrow Ictal$ \\ Break;}
    \uElseIf{$Score\,[\,Preictal\,] > \theta_P$}{$Pred\_Event \leftarrow Preictal$ \\ Break;}
    \Else{Continue;}
}
\end{algorithm}

The algorithm uses the DL model's segment-based classification results as inputs. The evaluation is based on a sliding window of $M$ segments. A score for the present window to be in an ictal phase is calculated based on two weighted terms: (i) how many individual segments generate a classification result of ictal phase, weighted by a coefficient of $\alpha_I$, and (ii) how many times the ictal classification repeats in a row, weighted by a coefficient of $\beta_I$. The score for the present window to be in a preictal phase is estimated in an equivalent way, based on the two terms corresponding to preictal phases, weighted by coefficients $\alpha_P$ and $\beta_P$. Finally, two pre-defined thresholds $\theta_I$ and $\theta_P$ are used to determine the final classification result from the sliding window. Once a score crosses the threshold, the algorithm will mark the event and break from evaluating the current sliding window. This is mainly to avoid delays during real-time seizure detection. A new evaluating sliding window starts immediately after the previous window terminates. Compared with traditional majority voting or moving average based algorithms, the proposed algorithm favors a prediction if the same classification results appear in a continuous manner. This significantly reduces the FPR in the classification. Moreover, the detection latency is not limited by the sliding window length as the algorithm terminates once the thresholds are crossed.

\subsection{Model Compression and Quantization}

For the models to be successfully deployed in a MCU, we applied compression and quantization techniques to reduce the computational cost. There are existing channel pruning techniques for reducing model dimension without retraining with data \cite{he2017channel}. However, since our model architectures are compact, the most effective method of compression is through iterations of retraining.

To select the most informative channels as the model inputs, we ranked all available recording channels based on the line length feature of these channels during the ictal phase. Line length is a measure of both high amplitude and high-frequency content of a time-domain signal, and is proven to be among the most effective features of seizures \cite{esteller2004comparison,logesparan2012optimal}. It is computed as:
\begin{equation}\label{eq_linelength}
  f_L(x_i) = \frac{1}{N}\sum_{t=1}^{N-1}|x_i(t-1) - x_i(t)|
\end{equation}
where $x_i(t)$ is the EEG signal of channel $i$ at time $t$, $N$ is the sample count of the ictal segment. The line length feature was only computed offline for channel ranking. It was not used in real-time seizure detection.
Based on the channel ranking, the top $K$ channels were used as the inputs for the models. The classification performances using different $K$ were compared for determining the optimal channel set. Similarly, we compared the performance using different data segment length $N$.

Computing a DL model using high arithmetic precision doesn't necessarily improve performance \cite{han2015deep}. Quantization of high precision coefficients not only saves the computational cost, but also reduces the memory cost for storing the coefficients. In fact, loading the coefficients from memory to the arithmetic unit can dominate the energy consumption in a microprocessor \cite{le2018mixed}. In this work, we quantized all coefficients to 8-bit fixed-point numbers. To evaluate the effects of model quantization, the performance of 8-bit quantized models was compared to that of 16-bit models (referred to hereafter as unquantized models). Together with the computationally efficient non-linear activation functions (Appendix Eqs. \ref{eq_relu}, \ref{eq_hard_sigmoid} and \ref{eq_softsign}) used in the model, the computational cost and latency were reduced.

\subsection{Training and Testing Methods}

The DL models were trained by supervised learning using the annotated data segments. Adam optimizer was used in training all of the models \cite{kingma2014adam}.
The experiments for model compression and quantization were performed in a 10-fold cross validation (CV). We used stratified CV where each fold has equal number of instances for each class. The number of segments depends on the available ictal onset time in each subject's recording, and the data segment length. The total data segments used during the 10-fold CV were less than 1\% of the whole recording per subject. 

After finalizing the model architectures, we evaluated the performances using a variant of the leave-one-out cross validation (LOOCV). For a subject's recording that contains $K$ seizure events, the recording is divided into $K$ sections with one seizure event in each of them. One section is used for validation and the rest $K-1$ sections are used for training. The process is repeated $K$ times so that all data is used exhaustively for validation. LOOCV suffers from large variation when $K$ is small \cite{kearns1999algorithmic}. In this work, we evaluate cases in the CHB-MIT that have at least 5 seizures in the recording, which lead to 15 cases including chb01, 03, 05, 06, 08, 10, 12-16, 18, 20, 23 and 24. Unlike the segment based 10-fold CV, the LOOCV is performed in a continuous manner in real-time using the sliding WMV algorithm.

Standard classification metrics were used, including accuracy, sensitivity, specificity, and FPR:

\begin{equation}\label{eq_sensitivity}
  Sensitivity = \frac{TP}{TP+FN}
\end{equation}

\begin{equation}\label{eq_specificity}
  Specificity = \frac{TN}{TN+FP}
\end{equation}

\begin{equation}\label{eq_accuracy}
  Accuracy = \frac{TP+TN}{TP+TN+FP+FN}
\end{equation}

\begin{equation}\label{eq_fpr}
  FPR \; (h^{-1}) = \frac{FP}{\textrm{Total recording length (in hours)}}
\end{equation}
where TP, TN, FP, and FN are true positive, true negative, false positive, and false negative detection, respectively. During the segment-based evaluation, results were directly compared with the annotated labels. For event-based seizure detection, we made the following definitions:
\begin{enumerate}
  \item If a detection event happens within 5 seconds of the seizure start time, it is considered as a TP. There can be at most one TP per genuine seizure.
  \item If no detection event happens within 5 seconds of the seizure start time, it is considered as a FN. There can be at most one FN per genuine seizure.
  \item If a detection event happens earlier than 5 seconds of the seizure start time, or 5 seconds after the seizure end time, it is considered to be a FP. The FP count is not limited by the number of seizures.
\end{enumerate}

Similarly, we made the definitions for event-based seizure prediction as follows:
\begin{enumerate}
  \item If a warning alert happens within 40 min before the seizure start time, it is considered as a TP. There can be at most one TP per genuine seizure.
  \item If no warning alert happens within 40 min before the seizure start time, it is considered as a FN. There can be at most one FN per genuine seizure.
  \item If a warning alert happens earlier than 40 min before or after the seizure start time, it is considered to be a FP. The FP count is not limited by the number of seizures.
\end{enumerate}

MATLAB\textsuperscript{\textregistered} was used for data handling. The DL models were implemented in Python with Tensorflow, which is an open-source machine learning library developed by the Google Brain team \cite{abadi2016tensorflow}. The DL training was performed on Google Cloud clusters using tensor processing units. 

\subsection{Hardware Implementation}

After training, the compressed and quantized models were deployed on a low-power MCU. We used a 32-bit ARM\textsuperscript{\textregistered} Cortex-M4 based MCU nRF52840 from Nordic Semiconductor \cite{semiconductor2019nrf52840}. The nRF52840 features a flash memory size of 1 MB and a RAM size of 256 kB. The MCU runs at a clock rate of 64MHz. The Cortex-M4 core supports multiple types of hardware multiplication in one clock cycle, including 16-bit signed multiplication with 32-bit results \cite{arm2009cortex}. The integrated floating-point unit was not used in this work.

The trained DL models were implemented in C/C++ for programming the MCU. The code was developed using Keil\textsuperscript{\textregistered} MDK \cite{keil2020mdk}. Open-source and commercial tools can be used to assist the code conversion from trained DL models to embedded systems \cite{matlab2020dltoolbox,google2020tfl}. The neural signal acquisition process was not implemented in this work. Instead, the integrated USB 2.0 full speed module was used for transferring the recorded EEG data from the computer to the MCU. The input data segments were buffered in the RAM. The direct memory access module was used for transferring the data so that the CPU was not interrupted \cite{liu2015pennbmbi}. The inference results were returned to the computer via the same USB interface.

\section{Results}

\subsection{Model Optimization}

\begin{figure}[!ht]
  \centering
  \includegraphics[width=0.65\textwidth,center]{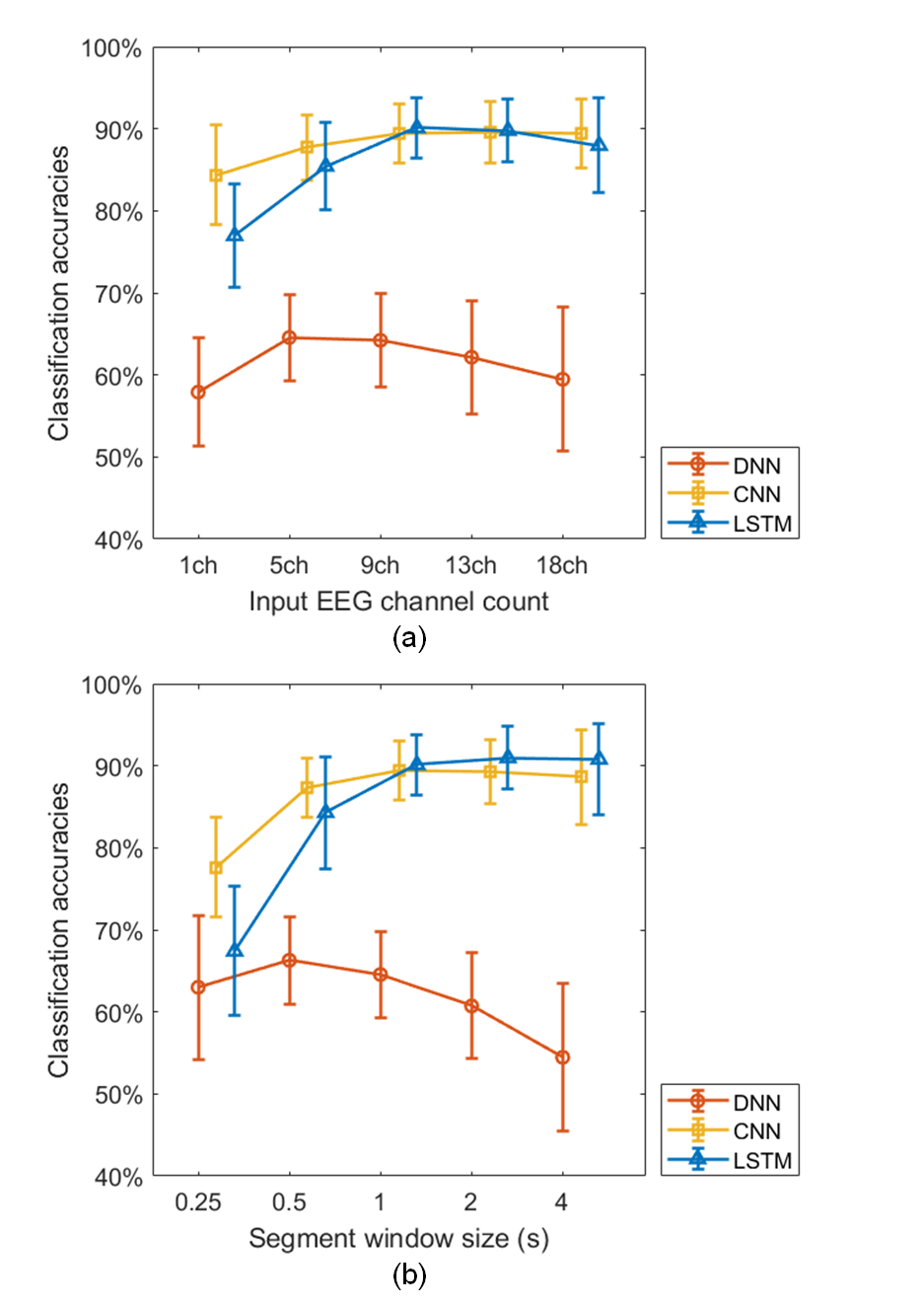}
  \caption{Model optimization during 10-fold CV. Each model's segment-based classification accuracy is plotted with error bars showing 95\% confidence interval. (a) Using 1, 5, 9, 13, or 18 EEG channels as the inputs. The channel selection was based on the line-length ranking algorithm. The segment length was 1 sec. (b) Using segment lengths of 0.25s, 0.5s, 1s, 2s, or 4s. The number of input channels for the DNN, CNN, and LSTM model was 5, 9, and 9, respectively. }\label{per_training}
\end{figure}

Three DL models were trained on all 24 cases (from 23 patients) in the CHB-MIT database. Models were not quantized during the training and optimization phase. The input channels were selected among all available EEG channels for each patient using the proposed line-length based ranking (Eq. \ref{eq_linelength}). Figure \ref{per_training} (a) shows the classification performance of each model using 1, 5, 9, 13, or 18 channels as the inputs. The segment length was chosen to be 1 sec (256 samples) in this analysis. The experimental result suggests that the optimal channel count is different for each model. Given the existing model size and experimental setup, the DNN model can take no more than 5 channels before it fails to extract critical information. More channels were generally helpful for the CNN model, but the performance improved marginally beyond 9 channels. The LSTM model reached peak performance at 9 and 13 channels, while more channels only caused additional variance.

Segment length is another key parameter that impacts the overall performance as well as the computational cost. Figure \ref{per_training} (b) shows the classification performance of each model using a segment length of 0.25 sec, 0.5 sec, 1 sec, 2 sec, or 4 sec. The ictal segments were generated with 50\% overlap in all cases. In the DNN model, using 5 channels as the input, the optimal performance was obtained with a segment length of 0.5 sec. In the CNN model, using 9 channels as the input, the optimal performance was obtained with a segment length of 1 sec. In the LSTM model, also using 9 channels as the input, the optimal performance was obtained with a segment length of 2 sec.

We targeted the optimal performance of each model that the hardware resources permit. In certain model configurations, using more channels as inputs may reduce the segment size for achieving similar performance. Depending on the system's requirement and the available hardware resources, one may prefer to use more input channels or a larger segment size. In practice, more input channels require more wearable/implantable electrodes and corresponding recording hardware, such as low-noise neural amplifiers. On the other hand, a larger window size requires more model coefficients thus more memory and RAM resources, and may cause a systematic latency. It should also be noted that increasing the segment length reduces the number of training and testing data segments, especially for patients with short ictal phases (eg. 6-9 sec in chb16). This could be a limiting factor in training DL models using a limited database. The hyperparameters used for training should be carefully tuned to minimize the impact.

\begin{figure}[!ht]
  \centering
  \includegraphics[width=0.7\textwidth]{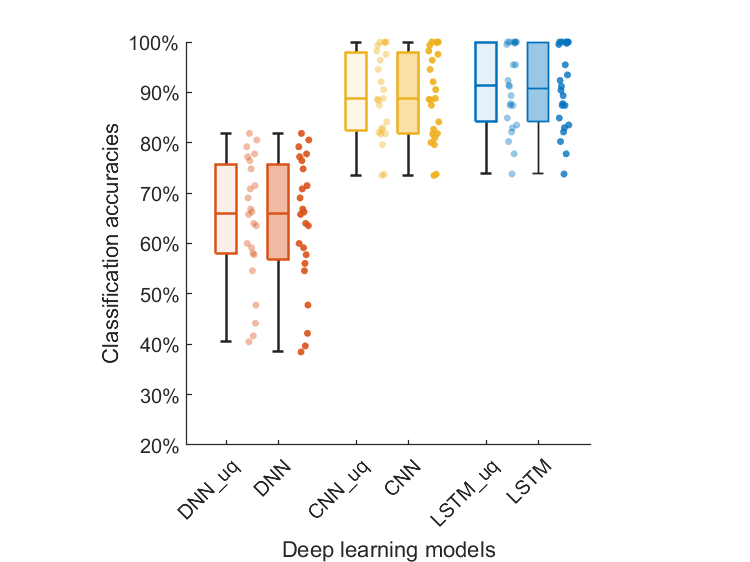}
  \caption{Performance of segment-based stage classification accuracy of each model. DNN\_uq, CNN\_uq, LSTM\_uq are the unquantized models. DNN, CNN, and LSTM are the quantized models. The box tops indicate 75th percentiles, box bottoms indicate 25th percentiles, solid lines indicate medians, whiskers indicate the span of the data, and dots show data points (from all 24 cases). }\label{per_testing_seg}
\end{figure}

Figure \ref{per_testing_seg} shows the segment-based classification accuracy of each model. The DNN\_uq, CNN\_uq, and LSTM\_uq indicated in the figure are the 16-bit unquantized models, while DNN, CNN, and LSTM are the 8-bit quantized versions. The average performance degradation due to quantization was less than 0.5\%. The LSTM model achieved the best overall performance (90.94\%), followed by the CNN model (89.21\%). The performance of the DNN model was relatively poor (64.55\%).

To compare the classification performance with non-DL algorithms, we constructed a linear discriminant analysis (LDA) classifier. Spectral amplitudes in selected frequency bands were used as the input features. The LDA classifier was chosen because it has been used in the Medtronic Activa PC+S device \cite{baldassano2019cloud}. The selected frequency bands were 0-2.7Hz, 2.7-5.4Hz, 5.4-10.8Hz, 10.8-21.7Hz, 21.7-43.4Hz, and 43.4-86.8Hz \cite{subasi2010eeg}. Table \ref{table_per_seg} shows the segment-based classification accuracy, sensitivity, and specificity of the LDA classifier and the three DL models for 10-fold CV. The DL models achieved better performance than the LDA classifier. Among the three DL models, the CNN and LSTM models showed superior overall performance, while the DNN model had a limited ability in discriminating the preictal and ictal phases.

\begin{table}[!ht]
\centering
\renewcommand{\arraystretch}{1.2}
\caption{Segment-based Classification in 10-fold CV using Quantized Models}\label{table_per_seg}
\vspace{0.3cm}
\footnotesize
\begin{tabular}{|c|c|c|c|c|c|}
\hline
\multicolumn{2}{|c|}{}                    & LDA  & DNN     & CNN     & LSTM    \\ \hline
\multicolumn{2}{|c|}{Overall Accuracy}    & 62.07\%  & 64.55\% & 89.21\% & 90.94\% \\ \hline
\multirow{4}{*}{Sensitivity} & Ictal      & 83.67\%  & 82.52\% & 96.59\% & 97.30\% \\ \cline{2-6}
                             & Preictal   & 49.24\%  & 51.83\% & 88.22\% & 91.46\% \\ \cline{2-6}
                             & Interictal & 55.53\%  & 60.66\% & 83.67\% & 85.33\% \\ \cline{2-6}
                             & Avg.       & 62.81\%  & 65.00\% & 89.50\% & 91.53\% \\ \hline
\multirow{4}{*}{Specificity} & Ictal      & 83.05\%  & 88.89\% & 97.79\% & 98.68\% \\ \cline{2-6}
                             & Preictal   & 73.55\%  & 75.29\% & 90.47\% & 91.58\% \\ \cline{2-6}
                             & Interictal & 74.94\%  & 83.75\% & 96.32\% & 97.00\% \\ \cline{2-6}
                             & Avg.       & 77.18\%  & 82.64\% & 94.86\% & 95.75\% \\ \hline
\end{tabular}
\end{table}

\subsection{Model Performance Evaluation}

\begin{figure}[!ht]
  \centering
  \includegraphics[width=1\textwidth]{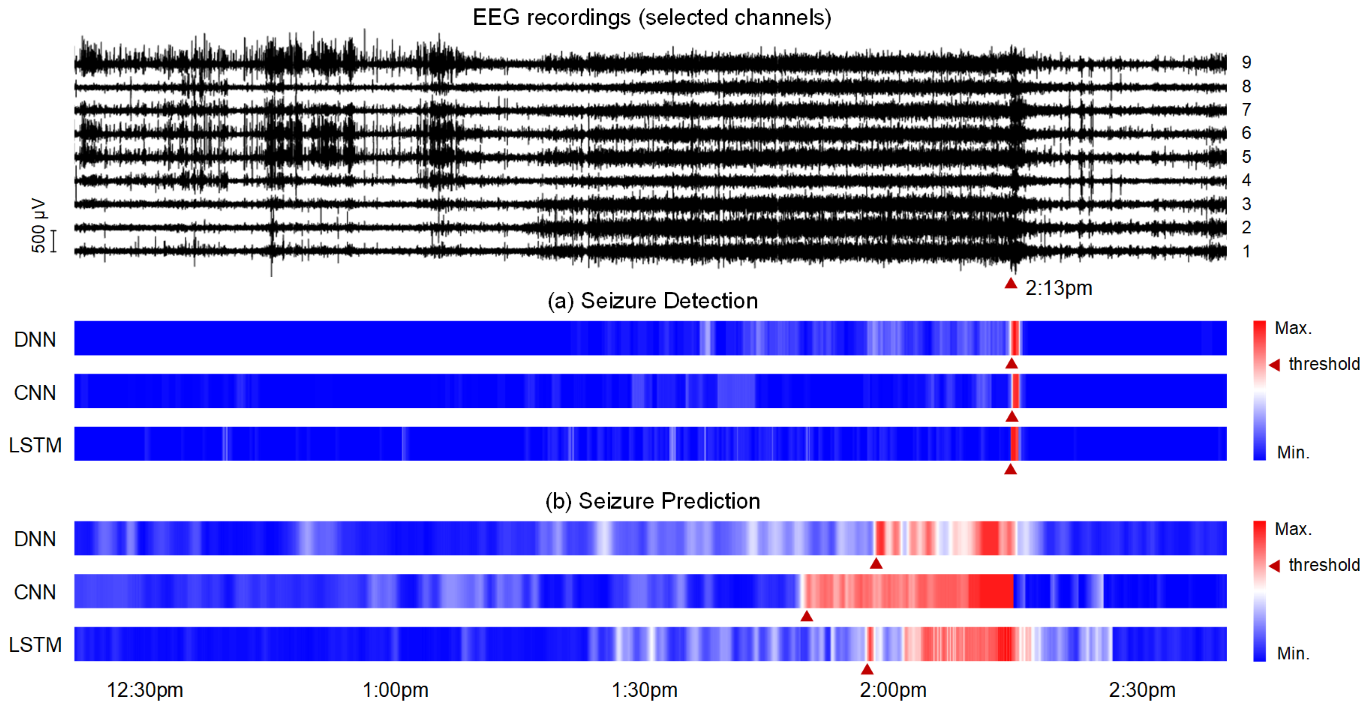}
  \caption{The detection and prediction of one seizure event of patient chb01 using the three DL models and sliding WMV algorithm. The genuine seizure occurred at 2:13 pm as annotated by clinical experts. Selected EEG channels used by the models are shown at the top (the DNN model used only the upper 5 channels). The scores of ictal (a) and preictal (b) phases calculated by the WMV algorithm are plotted. The normalized score is coded by color, with dark blue being the lowest and dark red being the highest. The earliest detection of each model is marked with a red arrow.}\label{per_example}
\end{figure}

Real-time seizure event detection and prediction using the trained, quantized DL models were simulated by streaming the selected EEG time series to the MCU in a continuous manner. Figure \ref{per_example} shows an illustrative example of the inferred scores of ictal ($Score\,[Ictal]$ in Algorithm \ref{alg_voting}) and preical ($Score\,[Preictal]$ in Algorithm \ref{alg_voting}) phases calculated by each model at each instance in time. The selected input EEG channels for the CNN and LSTM models are shown at the top of Figure \ref{per_example}. Only the upper 5 channels were used as inputs to the DNN model. All three models successfully detected the seizure event within a 5 sec window around the marked start time (Figure \ref{per_example} (a)). Furthermore, all three models successfully predicted the seizure within a 40 min horizon prior to the actual onset time (Figure \ref{per_example} (b)). In this example, the CNN model showed the best robustness against false positives in predicting seizures.

\begin{figure}[!ht]
  \centering
  \includegraphics[width=0.9\textwidth]{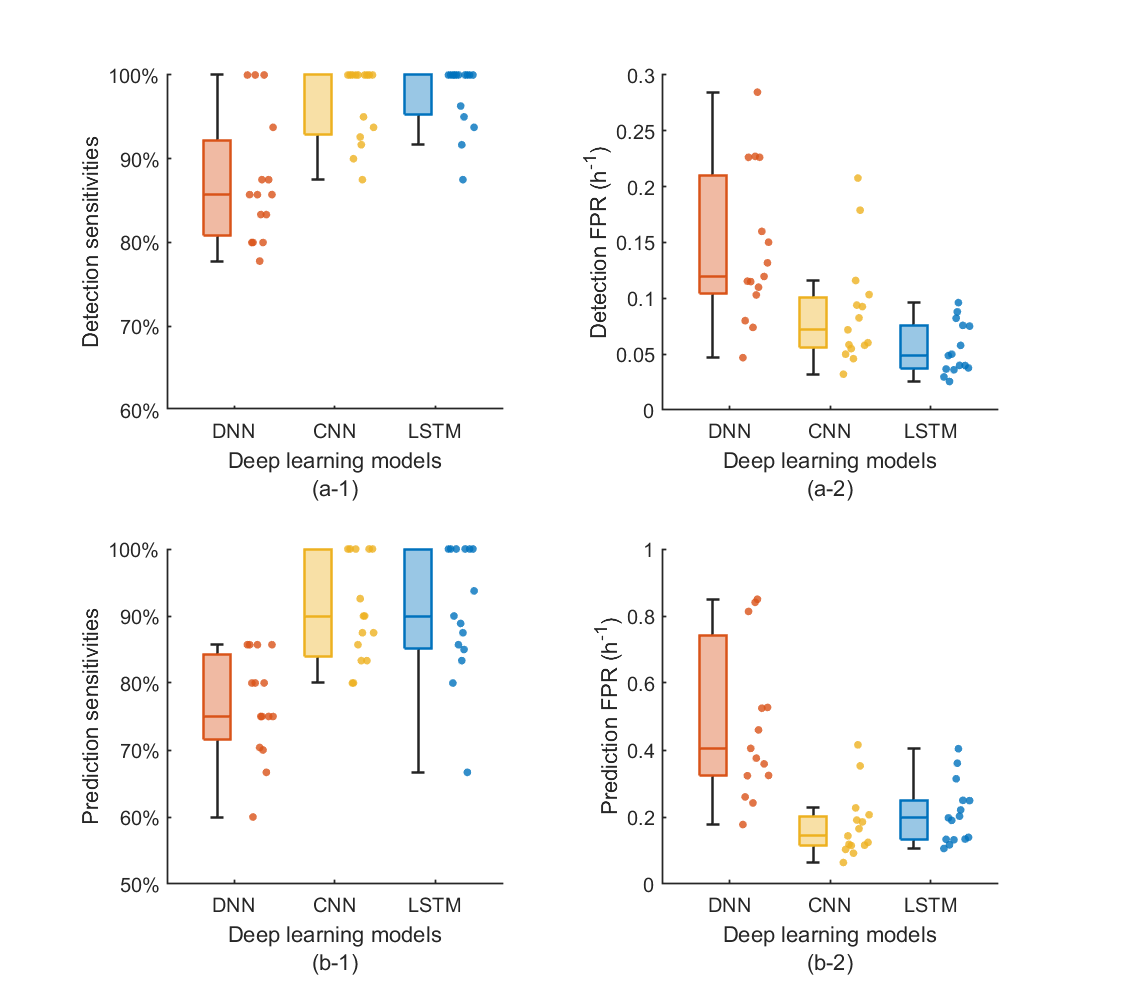}
  \caption{Performance of event-based seizure detection (a-1 \& a-2) and seizure prediction (b-1 \& b-2) in LOOCV.}\label{per_testing_event}
\end{figure}

The performance of event-based seizure detection and prediction was tested with LOOCV. The performance of the WMV algorithm was compared with a traditional moving average. The WMV algorithm improved the seizure detection FPR from 0.745$h^{-1}$ to 0.169$h^{-1}$ and the seizure prediction FPR from 2.341$h^{-1}$ to 0.710$h^{-1}$.
Figure \ref{per_testing_event} shows the box chart of each model's performance in seizure detection and prediction tasks in LOOCV. For seizure detection, the LSTM model achieved the highest average sensitivity (97.61\%) and the lowest FPR (0.071 $h^{-1}$), which corresponds to one false alarm in every 14.1 hours. For seizure prediction, the LSTM and CNN models achieved a comparable sensitivity above 90\%. The CNN model had a better average FPR (0.204 $h^{-1}$), which corresponds to one false alarm in every 4.9 hours.
The performance of the DNN model was relatively poor in seizure prediction tasks. The average and median sensitivities and FPRs of each model are summarized in Table \ref{table_per_event}.

\begin{table}[!ht]
\centering
\renewcommand{\arraystretch}{1.2}
\caption{Performance of Event-based Seizure Detection and Prediction in LOOCV}\label{table_per_event}
\vspace{0.3cm}
\footnotesize
\begin{tabular}{|c|c|c|c|c|c|c|c|c|}
\hline
\multirow{3}{*}{} & \multicolumn{4}{c|}{Detection}                                         & \multicolumn{4}{c|}{Prediction}                                        \\ \cline{2-9}
                  & \multicolumn{2}{c|}{Sensitivity} & \multicolumn{2}{c|}{FPR $(h^{-1})$} & \multicolumn{2}{c|}{Sensitivity} & \multicolumn{2}{c|}{FPR $(h^{-1})$} \\ \cline{2-9}
                  & Avg.            & Median         & Avg.             & Median           & Avg.            & Median         & Avg.             & Median           \\ \hline
DNN               & 87.36\% & 85.71\% & 0.169 & 0.140 & 76.66\% & 75.00\% & 0.710 & 0.474 \\ \hline
CNN               & 96.70\% & 100\%   & 0.102 & 0.084 & 90.66\% & 90.00\% & 0.204 & 0.168 \\ \hline
LSTM              & 97.61\% & 100\%   & 0.071 & 0.063 & 90.72\% & 90.00\% & 0.241 & 0.227 \\ \hline
\end{tabular}
\end{table}

\begin{figure}[!ht]
  \centering
  \includegraphics[width=1.2\textwidth,center]{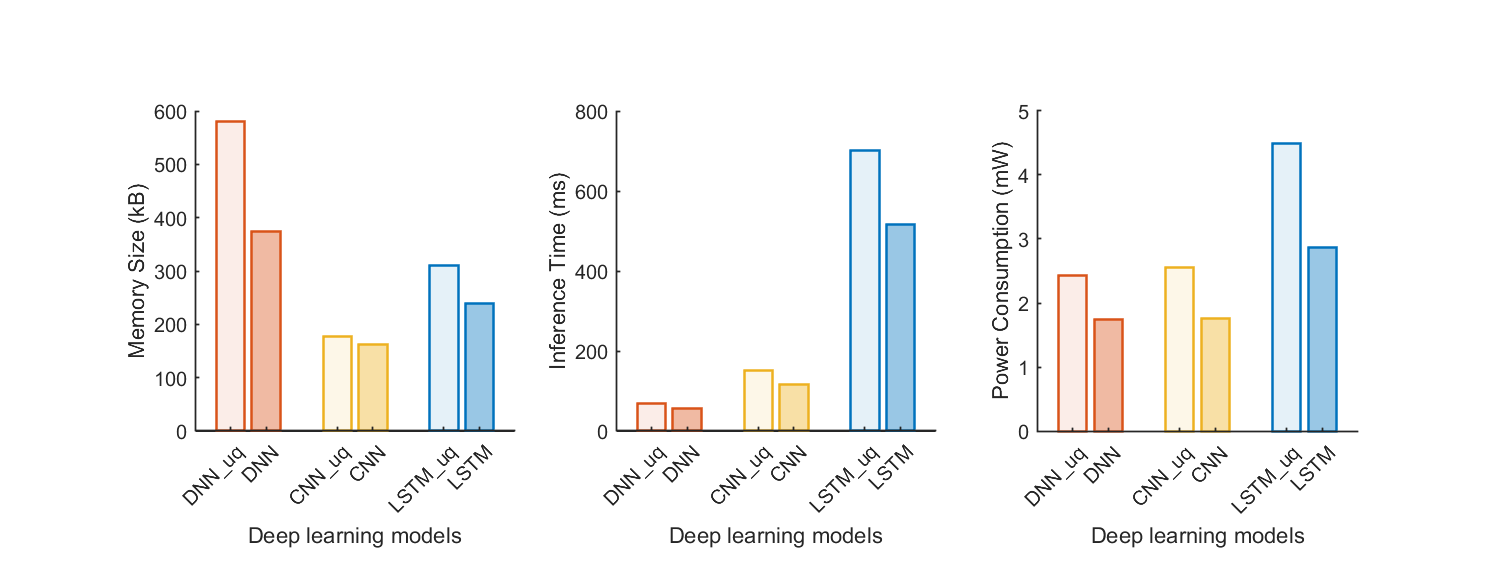}
  \caption{Each deep learning model's (a) memory size, (b) inference execution time, and (c) power consumption. DNN\_uq, CNN\_uq, LSTM\_uq are the unquantized models. DNN, CNN, and LSTM are the quantized models. }\label{per_metrics}
\end{figure}

Finally, each model's memory size, inference execution time, and power consumption are shown in Figure \ref{per_metrics}. The results before and after quantization were plotted for comparison. The memory size reflects the actual hardware implementation including the code overhead for data handling. The data transfer time via the USB port was excluded from the inference time since this would not be present in autonomous neural implants. The CPU core was put in sleep mode after executing the inference, and the power consumption was measured directly from the power supply. The reported power consumption is an average within the segment period. 
The DNN model required the largest memory size, while the CNN model required the least. This is mainly because of the limited convolutional kernel size used in the CNN model. The LSTM model had the longest inference time, but it was still within its segment period. From the perspective of power consumption, the DNN and CNN models were comparable, while the LSTM model consumed the most.

\section{Discussion}

Each of the three DL models developed in this work has its own strengths and weaknesses. The DNN model achieved the shortest inference time with minimum power consumption. The CNN model achieved a balanced performance with moderate power consumption and the smallest memory cost. The LSTM model achieved the best overall performance (e.g. highest seizure detection sensitivity and lowest FPR) at the expense of relatively long inference time and high power consumption. The optimal choice of model mainly depends on the specific application as well as the available hardware resources of the neural implant, including the batteries.

Model quantization improved inference time, memory size, and power consumption without sacrificing inference performance. On the other hand, model compression and channel pruning may require iterations of retraining. Layer size should be scaled with the sampling rate for capturing temporal features. Cascading DL models with a second stage sliding window-based algorithm, such as the sliding window based WMV algorithm implemented in this work, may compensate for the limitation of small input data buffer size of the edge DL models. Optimizing the design jointly for available hardware resources and real-time operational requirements is the key to achieving a satisfactory overall performance.

For event-based seizure detection, the CNN and LSTM models achieved an average high sensitivity of 96.7\% and 97.61\%, and low FPR at 0.102 and 0.071, respectively.
The results suggest that they are potential candidates for real-time, closed-loop therapeutic systems.
For seizure prediction, the performance of the DNN model was limited. We compare the seizure prediction performance of this work with prior studies (Table \ref{table_comparison}). Importantly, most of the listed studies assumed no time and computational resource limitations. The highest performances were achieved at the cost of sophisticated, hand-crafted, patient-specific feature engineering \cite{tsiouris2018long}. However, such complicated processing typically prevents its application in embedded devices for real-time, closed-loop inference. The relatively good hardware-based seizure prediction performance of our CNN and LSTM models compared to prior software implementations provides support for their use in future therapies.

One caveat to seizure prediction is that more advanced notice is not always better, unless a seizure prediction horizon (SPH), which is a seizure-free warning period between the alarm and the actual seizure onset, can be guaranteed by the algorithm \cite{winterhalder2003seizure}. Otherwise, an early warning may increase the anxiety of the patient given that a seizure may or may not happen any time within a long period after the warning. Including an accurate SPH significantly increases the difficulty in prediction \cite{winterhalder2003seizure}. It is not applicable for the light-weight DL models developed in this work. It should be noted, however, that the proposed edge DL inference paradigm doesn't preclude the functionality of uploading data to an external system including cloud-based computing resources for further analysis. Periodical diagnosis and model updating may be necessary for clinical adoption. But since these operations are not continuous during everyday use, data encryption can be reinforced and the peaking power dissipation during these short periods is not a big concern.

Another caveat to the results is that the EEG signals in the selected data set were recorded noninvasively using scalp electrodes. Therapeutic neural implant devices typically acquire EEG signals intracranially with a much higher signal-to-noise ratio (SNR) \cite{youngerman2019seeg}. The performance of DL algorithms typically improves with higher SNR signals, since more subtle neural features can be unveiled in these recordings \cite{truong2018convolutional}. We expect the performance of the models can be further improved, or the size and power consumption can be reduced if intracranial EEG recordings are used as the input.

Although this work uses a general-purpose, off-the-shelf MCU as the edge hardware platform, the design and optimization methods are applicable to ASIC development for clinical neural implants. The MCU core can be directly integrated into an ASIC design with the required memory or the models can be directly synthesized in the register-transfer level for minimizing the design overhead. Furthermore, in-memory or near memory computational techniques can be used to further reduce the power consumption of repeated multiply-accumulate (MAC) operations \cite{jia2020programmable,si2019twin}. An ASIC system-on-chip (SoC) that integrates analog neural interfaces (e.g. neural recorders and stimulators), digital DL inference module, power management and wireless communication modules can achieve the optimal power consumption and device footprint for chronic neural implants.

Finally, neural implants with different clinical purposes use various types of input neural signals, inference model complexities, and control strategies. Our work may serve as a reference for related studies on cognitive monitoring, sleep interventions, BMI, and other applications with closed-loop neuromodulation or neuroprosthetic control. The generalizability of our edge DL approach depends on the model complexity required to achieve high performance in these different applications. While DL model complexity has been defined in different ways, it is specifically the speed of inference, along with memory requirements, that is critical for resource-restrained edge devices \cite{Sze2017efficient}. To assess generalizability, we sought to compare our models' inference speed with that of a sample of prior DL studies in different neural application domains. Elapsed inference time (Fig. \ref{per_metrics} (b))  is rarely reported and is hardware dependent. A hardware-independent measure is the number of required computations (i.e. MAC operations) based on model architecture \cite{Taghavi2019hardware}. Restricting the comparison to CNN models, which are used in the majority of neural applications \cite{roy2019deep}, the total number of operations in the Conv and FC layers provide a reasonable estimate of complexity (see Appendix). Our CNN model required approximately 2.4M MACs to process the $256 \times 9$ input matrix. This complexity is higher than many prior CNN models used for BMI applications like P300 detection \cite{Cecotti2011convolutional}, steady-state evoked potential classification \cite{Kwak2017convolutional}, and attentive state detection \cite{Fahimi2019inter}, which ranged from about 0.1M to 0.5M MACs. Although CNN models for other tasks like sleep scoring \cite{tsinalis2016automatic} can be significantly more complex, on the order of 10M to 100M MACs, we suggest that edge DL systems are a realistic option for many BMI applications. Since DL eliminates the domain-specific manual feature selection used in conventional algorithms, research progress and technological advancement in one specific application area should readily generalize to other applications.

\begin{landscape}
\begin{table}[!ht]
\footnotesize
\renewcommand{\arraystretch}{1.2}
\caption{Comparison with Prior Seizure Prediction Studies}\label{table_comparison}
\vspace{0.2cm}
\begin{tabular}[t]{@{}lllllllll@{}}
\toprule
Year & Publication                                          & database & Features               & Algorithm  & Sensitivity & FPR (h$^{-1}$ )   & \begin{tabular}[c]{@{}l@{}}Prediction\\ Duration\end{tabular} \\ \midrule
2013 & Li \emph{et al.} \cite{li2013seizure} & Freiburg (21 cases) & spike rate & Threshold & 72.7\% & 0.11 & 50 min \\
2016 & Zhang \emph{et al.} \cite{zhang2016low}    & CHB-MIT (17 cases) & PSD ratio              & SVM       & 98.68\%     & 0.047 & 50 min         \\
2017 & Chu \emph{et al.} \cite{chu2017predicting} & CHB-MIT (13 cases) & Fourier Transform, PSD & Threshold & 83.33\%     & 0.392 & 86 min \\
2017 & Alotaiby \emph{et al.} \cite{alotaiby2017epileptic} & CHB-MIT (24 cases) & Spatial pattern statistics & LDA & 81\% & 0.47 & 60 min \\
2017 & Arabi \emph{et al.} \cite{aarabi2017seizure} & Freiburg (10 cases) & Uni-/bivariate features & Rule-based & 86.7\% & 0.126 & 30 min \\
2018 & Khan \emph{et al.} \cite{khan2017focal}    & CHB-MIT (15 cases) & Wavelet Transform      & CNN       & 87.80\%     & 0.142 & 10 min  \\
2018 & Truong \emph{et al.} \cite{truong2018convolutional}    & CHB-MIT (13 cases) & Short-time Fourier Transform      & CNN       & 81.20\%     & 0.16 & 30 min \\
2018 & Tsiouris \emph{et al.} \cite{tsiouris2018long}    & CHB-MIT (24 cases) & Wavelet Transform, PSD, statistics, etc. & LSTM       & 99.60\%     & 0.006 & 30 min \\
2018 & Shahbazi \emph{et al.} \cite{shahbazi2018generalizable} & CHB-MIT (14 cases) & Short-time Fourier Transform & LSTM & 98.2 \% & 0.13 & 45 min \\
2019 & Affes \emph{et al.} \cite{affes2019convolutional}    & CHB-MIT (24 cases) & Fourier Transform & C-GRNN  & 89.0\%     & 1.6 & 35 min \\
2020 & Abiyev \emph{et al.} \cite{abiyev2020identification} & CHB-MIT (7 cases) & Raw EEG & CNN & 97.67\% &    1.97 & N/A \\
 & This work & CHB-MIT (15 cases) & Raw EEG & \begin{tabular}[c]{@{}l@{}}DNN\\ CNN\\ LSTM\end{tabular}
 & \begin{tabular}[c]{@{}l@{}}76.66\% \\ 90.66\%  \\ 90.72\% \end{tabular}
 & \begin{tabular}[c]{@{}l@{}}0.710 \\ 0.204 \\ 0.241 \end{tabular}
 & \begin{tabular}[c]{@{}l@{}}40 min\\ 40 min\\ 40 min\end{tabular} \\
\bottomrule
\end{tabular}
\end{table}

\end{landscape}

\section{Conclusion}

In this work, we developed edge DL models and investigated their potential utility for future clinical applications involving neural implants. We adopted three commonly used DL architectures (DNN, CNN, and LSTM) and optimized the models for deployment in resource-restrained hardware. Using the CHB-MIT database, we show that edge DL inference can achieve comparable performance in epileptic seizure detection to many prior implementations that had no time and computational resource limitations. Our results suggest that edge DL inference is a promising option for closed-loop neuromodulation, with superior robustness and security compared to wireless communication-based solutions. While clinical studies are needed to confirm the efficacy of this paradigm, this work demonstrates the feasibility and potential advantages. We envision that the next generation of clinical neural implants could leverage edge DL inference to greatly improve their therapeutic benefit.

\section{APPENDIX}
\subsection{Processing and activation functions}
In the FC layers, the processing function of each neuron $j$ is given by:
\begin{equation}\label{eq_neuron}
  y_j = f_R(\sum_{i=1}^{n}w_{j,i}\cdot x_i + b_j)
\end{equation}
where $x_i$ is from the previous layer, $w_{j,i}$ is the weight factor, $b_j$ is the bias term, $n$ is the number of neurons in the previous layer, and $f_R(x)$ is the activation function. The $f_R(x)$ used in this work is the rectified linear unit (ReLU) function. ReLU is a piecewise linear function as given by:
 \begin{equation}\label{eq_relu}
   f_R(x) = max\{0, x\}
 \end{equation}

The processing functions of each LSTM cell $j$ are given by:
\begin{equation}\label{eq_ft}
  f_j = f_G(W_{h,f,j} \cdot h_{j-1} + W_{x,f,j} \cdot  x_j + b_{f,j})
\end{equation}
\begin{equation}\label{eq_it}
  i_j = f_G(W_{h,i,j} \cdot h_{j-1} + W_{x,i,j} \cdot x_j + b_{i,j})
\end{equation}
\begin{equation}\label{eq_ot}
  o_j = f_G(W_{h,o,j} \cdot h_{j-1} + W_{x,o,j} \cdot x_j + b_{o,j})
\end{equation}
\begin{equation}\label{eq_ctb}
  \tilde{c_j} = f_S(W_{h,c,j} \cdot h_{j-1} + W_{x,c,j} \cdot x_j + b_{c,j})
\end{equation}
\begin{equation}\label{eq_ct}
  c_j = f_j \cdot  c_{j-1} + i_j \cdot  \tilde{c_j}
\end{equation}
\begin{equation}\label{eq_ht}
  h_j = o_j  \cdot f_S(c_j)
\end{equation}
where $x_j$ is the output from the previous layer, $h_j$ is the hidden state, $c_{j}$ is the memory state ($\tilde{c_{j}}$ is the candidate), $i_j$ is the input gate, $o_j$ is the output gate, and $f_j$ is the forget gate. $W$'s and $b$'s are the weight and bias terms for each neuron in the LSTM cell.

The gate activation function $f_G(x)$ is given by:
\begin{equation}\label{eq_hard_sigmoid}
  f_G(x) = max\{0,min\{1,\frac{x}{5} + \frac{1}{2}\}\}
\end{equation}

The state activation function $f_S(x)$ is given by: 
\begin{equation}\label{eq_softsign}
  f_S(x) = \frac{x}{1+|x|}
\end{equation}

\subsection{Multiply-accumulate operation estimates}
The number of MAC operations for a CNN can be estimated as follows. For FC layers:
\begin{equation}
\#MAC_{fc} = C_{in}*C_{out}
\end{equation}
where $C_{in}$ is the number of input channels (neurons) and $C_{out}$ is the number of output channels. For Conv layers:
\begin{equation}
\#MAC_{conv} = C_{in}*m_{out}*n_{out}*C_{out}*h*v
\end{equation}
where $m_{out}*n_{out}$ is the size of the output feature map, $C_{out}$ is again the number of output channels (equivalent to the number of kernels), and $h*v$ is the size of each kernel. Max pooling and activation operations are ignored as they typically contribute very little to the total MAC count.

\section*{References}
\bibliography{iopart-num}

\end{document}